\documentclass[reprint,pre,groupedaddress,amsmath,longbibliography]{revtex4-2}
\usepackage{graphicx,txfonts,color}
\usepackage[colorlinks,linkcolor=blue,citecolor=blue,anchorcolor=blue,urlcolor=blue]{hyperref}

\begin{document}

\title{Emergence of biconnected clusters in explosive percolation}

\author{Liwenying Yang}
\affiliation{School of Physics, Hefei University of Technology, Hefei, Anhui 230009, China}

\author{Ming Li}
\email{lim@hfut.edu.cn}
\affiliation{School of Physics, Hefei University of Technology, Hefei, Anhui 230009, China}

\date{\today}

\begin{abstract}
By introducing a simple competition mechanism for bond insertion in random graphs, explosive percolation exhibits a sharp phase transition with rich critical phenomena. We investigate high-order connectivity in explosive percolation using an event-based ensemble, focusing on biconnected clusters, where any two sites are connected by at least two independent paths. Our numerical analysis confirms that explosive percolation with different intra-cluster bond competition rules shares the same percolation threshold and universality, with biconnected clusters percolating simultaneously with simply connected clusters. However, the volume fractal dimension $d_{f}'$ of biconnected clusters varies depending on the competition rules of intra-cluster bonds. The size distribution of biconnected clusters exhibits double-scaling behavior: large clusters follow the standard Fisher exponent derived from the hyperscaling relation $\tau'=1+1/d_{f}'$, while small clusters display a modified Fisher exponent $\tau_0<\tau'$. These findings provide insights into the intricate nature of connectivity in explosive percolation.
\end{abstract}

\maketitle

\section{Introduction}

Explosive percolation (EP), proposed by Achlioptas et al.~\cite{Achlioptas2009}, has become a prominent topic in percolation theory and network science~\cite{Boccaletti2016,DSouza2019,Li2021}. The core mechanism of EP is the suppression of large cluster growth when new bonds are inserted, a process known as the Achlioptas process~\cite{Achlioptas2009}. A typical example is the product rule~\cite{Achlioptas2009}: starting with a null graph, two potential bonds are chosen at each time step, and the bond that minimizes the product of the sizes of the clusters it connects is inserted, while the other is discarded. This can be generalized to the best-of-$m$ rule or min-cluster-$m$ rule~\cite{Friedman2009}, where more than two potential bonds are considered, and various criteria can be used for bond selection~\cite{Boccaletti2016}.

The most intriguing finding in EP models is the abrupt, first-order-like percolation transition. Although later studies confirmed that this transition is actually continuous~\cite{Costa2010,Lee2011,Grassberger2011,Riordan2011}, the intense scientific debate it sparked has significantly advanced percolation theory and network science~\cite{Boccaletti2016,DSouza2019,Li2021}. Methods developed to verify the discontinuity of EP, such as gap scaling~\cite{Manna2011,Nagler2011} and cluster-size heterogeneity~\cite{Lee2011}, have been applied to characterize the critical behavior of various systems~\cite{Noh2011,Lv2012,Jo2012,Fan2020,Feshanjerdi2021,Feshanjerdi2023}. The mechanisms underlying explosive phenomena have also proven useful in network structure analysis~\cite{Pan2011,Schroeder2018,Qiu2021} and immunization strategies~\cite{Clusella2016}.

Despite EP being widely recognized as a continuous phase transition, numerous studies have reported anomalous finite-size behaviors that deviate from standard finite-size scaling (FSS) theory~\cite{Grassberger2011,Friedman2009,Hooyberghs2011,Nagler2011,DSouza2015,Fan2020}. A recent study introduced a dynamic ensemble called the event-based ensemble~\cite{Li2023}, where EP adheres to standard FSS theory. This approach explains the anomalous finite-size behaviors observed in conventional ensembles with fixed bond density as a result of multiplex scalings induced by large fluctuations of the pseudo-critical point, where clean FSS can be consistently observed.

\begin{figure}
\centering
\includegraphics[width=0.8\columnwidth]{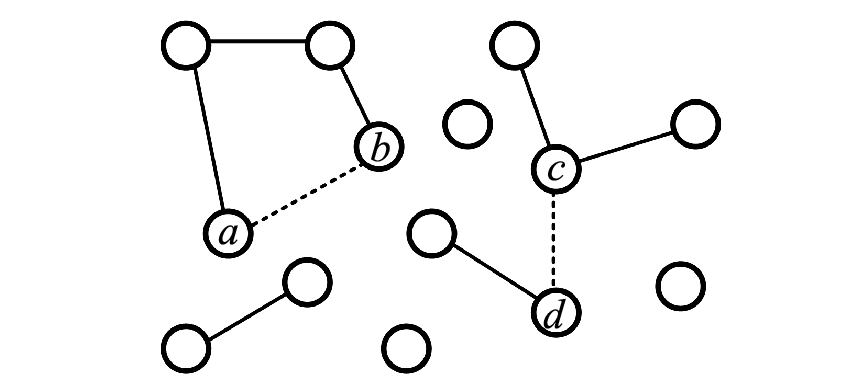}
\caption{A sketch of bond insertion rule in Achlioptas process, particularly for the differences between inter- and intra-cluster bonds. Here, circles, solid lines, and dashed lines represent sites, existing bonds, and potential bonds, respectively. A bond $e_{ab}$ whose ends $a$ and $b$ are in the same cluster is called intra-cluster bond, while a bond $e_{cd}$ whose ends $c$ and $d$ are in different clusters is called inter-cluster bond. The size $s$ of a cluster is the number of sites in it. For sites $a$, $b$, $c$, and $d$, the sizes of corresponding clusters are $s_a=s_b=4$, $s_c=3$, and $s_d=2$. For the inter-cluster bond $e_{cd}$, the size product is $P_{cd}=s_c \times s_d=6$. For the intra-cluster bond $e_{ab}$, there are three typical definitions of the size product: $P_{ab}=s_a\times s_b =16$, $P_{ab}=s_a=s_b=4$, and $P_{ab} = 0$, which we refer to as square, linear, and zero rules, respectively. Consequently, $e_{cd}$ is inserted when square rule is applied, and $e_{ab}$ is inserted for linear and zero rules.} \label{f1}
\end{figure}

The study of EP still faces unresolved issues and contradictions, particularly regarding the impact of intra-cluster bond insertion on critical phenomena. On one hand, inserting an intra-cluster bond, where both ends belong to the same cluster, does not directly affect cluster sizes but increases bond density, as depicted in Fig.~\ref{f1}. Thus, if potential bonds include intra-cluster ones, these might be preferentially inserted to curb cluster growth, aligning with the core mechanism of the Achlioptas process. On the other hand, adhering strictly to the product rule involves calculating size products without regard to whether the bond is intra-cluster, leading to a preference for inserting bonds between smaller clusters. These differences in bond insertion are illustrated in Fig.~\ref{f1} by visually defining the size product $P$ of an intra-cluster bond in a cluster of size $s$ as $P=s^2$, $P=s$, or $P=0$, which we refer to as square, linear, and zero rules, respectively.

\begin{figure}
\centering
\includegraphics[width=\columnwidth]{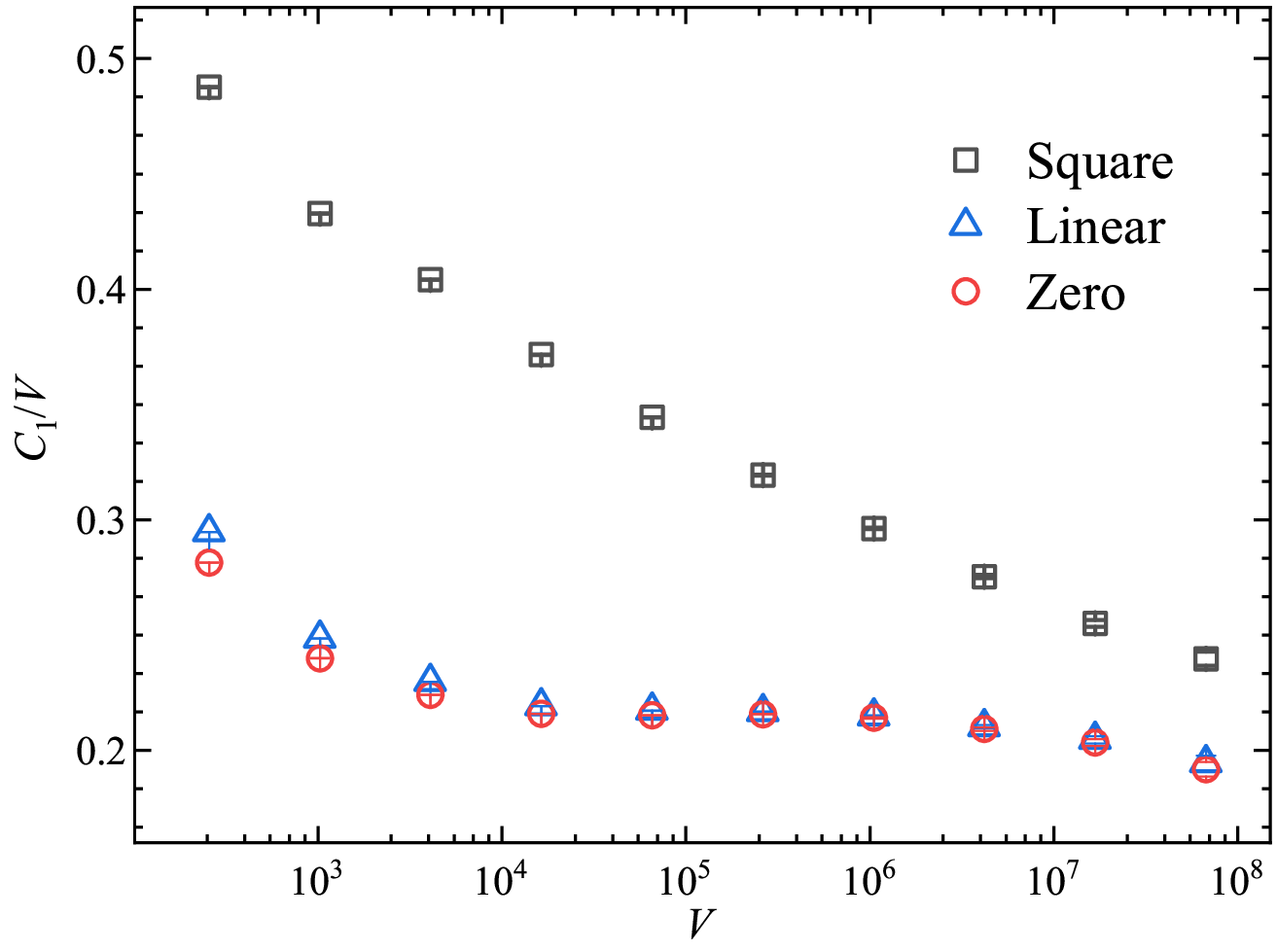}
\caption{The order parameter $C_1/V$ at the infinite-volume critical point $T_c$ is plotted as a function of the system volume $V$ for EP under product rule. Here, $C_1$ represents the size of the largest cluster. For square rule, the order parameter $C_1/V$ decreases as the system volume increases, and seems likely to vanish for infinite systems. Conversely, for linear and zero rules, where the insertion of intra-cluster bonds is prioritized, $C_1/V$ appears to approach a constant value for infinite systems. In simulations, the critical point is set to $T_c=0.8884491$~\cite{Li2023}, which is defined as the total number of bonds at the percolation threshold normalized by the system volume.}
\label{f2}
\end{figure}

In Fig.~\ref{f2}, we illustrate the FSS of the critical order parameter $C_1/V$ for various competition rules of intra-cluster bonds, where $C_1$ is the size of the largest cluster and $V$ is the total number of sites. It appears that linear and zero rules yield an asymptotic behavior of $C_1/V$ for increasing $V$, which was interpreted as an indication of a discontinuous percolation transition~\cite{Cho2011}. Conversely, employing the product rule without differentiation between intra-cluster and inter-cluster bonds, i.e., square rule, results in a vanishing order parameter in an infinite system, akin to standard percolation.

Note that at criticality, the probability that a randomly chosen bond is an intra-cluster bond vanishes as the system volume $V$ increases. This probability can be estimated as
\begin{equation}
V^{-1}\sum_{s=1}^{C_1} s^2n(s,V) \sim V^{d_f(3-\tau)-1}  \sim V^{-2(1-d_f)}.  \label{eq-pitb}
\end{equation}
Here, $n(s,V)$ is the cluster number density, defined as the number of clusters with size $s$ normalized by the system volume. Therefore, $s^2n(s,V)/V$ represents the probability that two randomly chosen sites belong to the same cluster of size $s$. At criticality, the cluster number density scales as $n(s,V)\sim s^{-\tau}$, where $\tau$ is the Fisher exponent. In random graphs, where there is no concept of side length, the fractal dimension $d_f$ is defined by the system volume relationship $C_1\sim V^{d_f}$, known as the volume fractal dimension, and the hyperscaling relation reduces to $\tau=1+1/d_f$, which is used in Eq.~(\ref{eq-pitb}). Since $d_f<1$, it is evident that the probability given by Eq.~(\ref{eq-pitb}) vanishes for large $V$. This indicates that no intra-cluster bonds can be chosen as potential bonds in the infinite-volume limit, so that, the numerical results of Fig.~\ref{f2}, which are similarly presented in Ref.~\cite{Cho2011}, cannot be used as effective evidence that systems of different competition mechanisms for intra-cluster bonds have different infinite-volume critical behaviors.

Furthermore, the insertion of intra-cluster bonds is crucial for the organization of high-order structures. A typical example of such structures is biconnected cluster (BC), where sites are connected by at least two independent paths. In standard percolation on random graphs, BC percolates at the same threshold as a connected cluster (CC) but exhibits different fractal dimensions~\cite{Newman2008,Huang2018}. Moreover, on low-dimensional hypercubic lattices~\cite{Xu2014,Zhou2015,Huang2018} and complex networks~\cite{Cellai2011,Liu2012,Li2015,Benson2016,Tian2017,Osat2020}, the non-trivial organization of high-order connectivity has unveiled significant geometric properties of percolation systems that cannot be captured by simple connections alone. For example, structures like the $k$-core, consisting of compact clusters, can exhibit hybrid transitions involving both a jump of the giant cluster and a critical singularity at the percolation threshold~\cite{Goltsev2006,Cellai2011,Osat2020,Gao2024,Shang2019,shang2020,shang2020a}. Additionally, the critical behaviors of high-order structures of percolation clusters have also been demonstrated by the so-called backbones~\cite{Gefen1981,Porto1997,Barthelemy1999,Fang2022}.

In this paper, our focus lies on exploring the high-order organization of critical clusters in EP. We reveal that EP of different competition rules for intra-cluster bonds share the same percolation threshold and universality, while the fractal dimension of BCs is rule-dependent. Additionally, the cluster number density of BCs shows a double-scaling behavior, also depending on the competition rule of intra-cluster bonds.

The remainder of this paper is organized as follows. Section~\ref{sec-ma} shows the details of the model, algorithm, and observables. In Sec.~\ref{sec-cc}, we show the simulation results of the EP under three competition rules of intra-cluster bonds. The FSS behaviors of BCs are studied in Sec.~\ref{sec-bc}. We include a short discussion in the last section.

\section{Model, Algorithm, and Observables}   \label{sec-ma}

The Achlioptas process initiates with a null graph of volume $V$, then proceeds by inserting bonds step by step. At each time step, two potential bonds, denoted as $e_{ab}$ and $e_{cd}$, are randomly selected from all unconnected pairs of sites. Here, $a$, $b$, $c$, and $d$ denote the sites. Subsequently, the size products $P_{ab}=s_a \times s_b$ and $P_{cd}=s_c \times s_d$ are computed, where $s_i$ denotes the size of the cluster that site $i$ belongs to. If $P_{ab}<P_{cd}$, bond $e_{ab}$ is inserted, and bond $e_{cd}$ is discarded. In the case of $P_{ab}=P_{cd}$, one bond is randomly chosen for insertion. This mechanism defines the product rule of EP~\cite{Achlioptas2009}.

For intra-cluster bonds, sites at the two ends belong to the same cluster (see Fig.~\ref{f1}), thus, a specialized definition for the size product $P$ is necessary. Generally, three approaches are considered: $P=s^2$, $P=s$, and $P=0$, where $s$ represents the size of the cluster to which the intra-cluster bond belongs. For convenience, we refer to them as square, linear, and zero rules, respectively. In zero rule ($P=0$), intra-cluster bonds are prioritized for insertion, leading to the system tending to form large dense clusters. Conversely, with square rule ($P=s^2$), intra-cluster bonds are hardly inserted into large clusters, due to their significantly larger size product. The linear rule ($P=s$) represents an intermediate scenario between these two scenarios.

To apply the event-based ensemble effectively, we need to define a dynamic pseudo-critical point for each individual realization, where all quantities are sampled and averaged. We propose two such critical points based on the sizes of the largest CC and BC, respectively. In each run of the Achlioptas process, we monitor the size of the largest CC, denoted as $\mathcal{C}_1(t)$, at each time step $t$. Then, we calculate the one-step incremental size of the largest CC as $\Delta(t)=\mathcal{C}_1(t+1)-\mathcal{C}_1(t)$. The dynamic pseudo-critical point $\mathcal{T}_V$ for a single realization is defined as $\mathcal{T}_V=t_{\text{max}}/V$, where $t_{\text{max}}$ represents the time step at which $\Delta(t)$ reaches its maximum value. Similarly, another dynamic pseudo-critical point $\mathcal{T}_V'$ can be defined based on the one-step incremental size of the largest BC, denoted as $\Delta'(t)=\mathcal{B}_1(t+1)-\mathcal{B}_1(t)$, where $\mathcal{B}_1$ represents the size of the largest BC.

In our simulations, we employ the Newman-Ziff algorithm to track the growth of $\mathcal{C}_1$ during the bond-insertion process~\cite{Newman2000}, as it allows for real-time updates of evolving clusters using a data structure known as disjoint set~\cite{Galil1991}. However, it is worth noting that a site can belong to multiple BCs simultaneously. Therefore, the Newman-Ziff algorithm, which relies on disjoint sets, is not suitable for storing information about BCs. To maintain dynamic BCs, we employ a data structure, called block forest~\cite{Westbrook1992}. The block tree is constructed by identifying blocks (BCs) and their articulation points, resulting in a tree structure where each node represents either a BC or an articulation point. An articulation point is a site whose removal increases the number of CCs in the graph, and edges in this tree represent the inclusion of articulation points within these BCs. When a newly inserted bond bridges two nodes (BCs or articulation points), the block tree is updated by condensing a chain of nodes between the two nodes into a new node (BC). This dynamic update allows for real-time recording of BC information. Hence, this data structure can be readily adapted to the EP model, enabling real-time tracking of the one-step incremental size of the largest BC.

In each realization of EP, we firstly identify the pseudo-critical points $\mathcal{T}_{V}$ and $\mathcal{T}_{V}'$, then at the two dynamic pseudo-critical points, we sample and calculate the following observables:
\begin{itemize}
\item The mean pseudo-critical point $T_{V}\equiv\langle\mathcal{T}_{V}\rangle$ and $T_{V}'\equiv\langle\mathcal{T}_{V}'\rangle$, and their fluctuations $\sigma(\mathcal{T}_V)\equiv\sqrt{\left\langle \mathcal{T}_{V}^2\right\rangle-\left\langle\mathcal{T}_{V}\right\rangle^2}$ and $\sigma(\mathcal{T}_{V}')\equiv\sqrt{\left\langle \mathcal{T}_{V}'^2\right\rangle-\left\langle\mathcal{T}_{V}'\right\rangle^2}$.
\item The size of the $n$-th largest CC, $C_{n}\equiv\langle{\mathcal C}_{n}\rangle$, and the size of the $n$-th largest BC, $B_{n}\equiv\langle{\mathcal B}_{n}\rangle$, where $\mathcal{C}_n$ and $\mathcal{B}_n$ refer to the values in a single realization.
\item The cluster number density of BCs, $n(s,V)\equiv\left\langle \mathcal{N}_s\right\rangle/V$, where $\mathcal{N}_s$ is the number of BCs with size $s$ in a single realization.
\end{itemize}
Here, the brackets $\langle \cdot\rangle$ denote the average of different realizations in the event-based ensemble.

\section{Percolation of connected clusters}    \label{sec-cc}

In this section, we study the percolation of CCs under different competition rules of intra-cluster bonds. The data shown in this section is extracted at the dynamic pseudo-critical point $\mathcal{T}_{V}$ identified by the largest one-step increment size of the largest CC.

\begin{figure}
\centering
\includegraphics[width=\columnwidth]{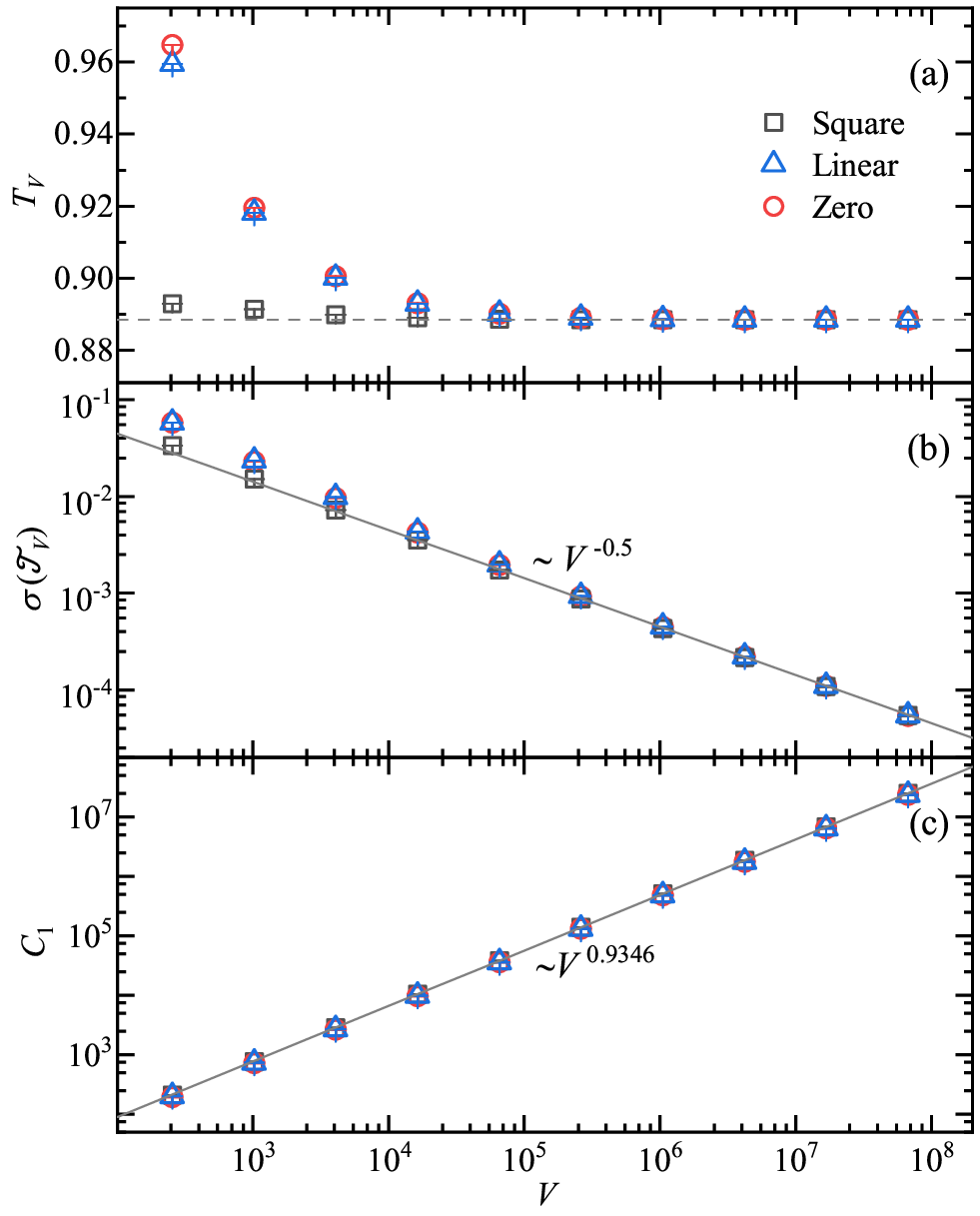}
\caption{The FSS of EP for different competition rules of intra-cluster bonds. (a) The asymptotic behavior of the pseudo-critical point $T_V$ versus system volume $V$. Although for finite $V$, the three scenarios show different pseudo-critical points $T_V$, they are convergent to the same value $T_c \approx 0.888449$ for large systems indicated by the dashed line. The fit results for these asymptotic behaviors are listed in Tab.~\ref{t1}. (b) The plots of the dynamic pseudo-critical point fluctuation $\sigma(\mathcal{T}_V)$ versus system volume $V$. The fit results in Tab.~\ref{t1} suggest that all three scenarios have the same FSS $\sigma \sim V^{-1/2}$ indicated by the solid line. (c) The size of the largest CC sampled at the dynamic pseudo-critical point, $C_{1}\equiv\langle{\mathcal C}_{1}\rangle$, as a function of system volume $V$. The solid line represents the fit result $d_f=0.9346$ in Tab.~\ref{t1}.}     \label{f3}
\end{figure}

\begin{table}[b]
\caption{The fit results of the infinite-volume critical point $T_c$, the reciprocal value of the correlation-length exponent $1/\nu$, the exponent $\theta$ to describe the fluctuation of $\mathcal{T}_{V}$, and the fractal dimension $d_f$ of the largest CC for different competition rules of intra-cluster bonds. Here, the data are all sampled at the dynamic pseudo-critical point $\mathcal{T}_{V}$, where the one-step increment size of $\mathcal{C}_1$ reaches its maximum value. Within error bars, the three scenarios suggest the same percolation threshold, and critical exponents.} \label{t1}
\begin{ruledtabular}
\begin{tabular}{llllll}
\text{Rule}   &  \multicolumn{1}{c}{$T_c$}  &  \multicolumn{1}{c}{$1/\nu$} &  \multicolumn{1}{c}{$\theta$} & \multicolumn{1}{c}{$d_f$}  \\
\hline
\text{Square} & 0.888\,449\,3(3)   & 0.74(5)   & 0.499(1)   & 0.9346(2)  \\
\text{Linear} & 0.888\,449\,3(4)   & 0.741(6)  & 0.495(5)   & 0.9347(1)  \\
\text{Zero}   & 0.888\,449\,0(2)   & 0.741(1)  & 0.49(2)    & 0.9346(3)
\end{tabular}
\end{ruledtabular}
\end{table}

In Fig.~\ref{f3} (a), we observe the asymptotic behavior of the pseudo-critical point $T_V$ plotted against the system volume $V$. For zero rule, the system exhibits the largest pseudo-critical point among the three scenarios, due to the preferential insertion of intra-cluster bonds, which decays the onset of percolation. Conversely, EP under square rule demonstrates the smallest pseudo-critical point due to its blind bond insertion approach, where intra-cluster bonds are not given special preference. Nevertheless, with the increasing of system volumes, these pseudo-critical points approach the same value, as indicated by the dashed line in Fig.~\ref{f3} (a).

To quantify these asymptotic behaviors, we fit the Monte Carlo data of $T_V$ to the FSS ansatz
\begin{equation}
T_V=T_c+V^{-1/\nu}(a_0+a_1V^{-\omega}+\cdots).  \label{Eq:PowerC}
\end{equation}
Here, $T_c$ denotes the infinite-volume critical point, $\nu$ is the critical exponent of the correlation length, and the term $V^{-\omega}$ is a correction to the FSS. If the correction term is excluded, i.e., $(a_1=0)$, the fitting results are sensitive to the changes of the lower cut-off $V_\text{min}$ on the data points admitted in the fit. With all the terms of the FSS ansatz Eq.~(\ref{Eq:PowerC}) free, we estimate the stable fit for $T_c$ and $1/\nu$ as listed in Tab.~\ref{t1}. Choosing the optimal fit for the FSS ansatz typically involves identifying the smallest value of $V_\text{min}$ for which the $\chi^2$ per degree of freedom is close to unity. Specifically, $\chi^2$ is calculated as the sum of the squared differences between observed values and the fitting curve, each normalized by the errors of observations. Moreover, further increases in $V_\text{min}$ should not result in significant reductions in the $\chi^2$ value beyond one unit per degree of freedom.

The results presented in Tab.~\ref{t1} demonstrate that, within the error margins, the pseudo-critical points for all three scenarios converge to the same percolation threshold, $T_c=0.888449$, aligning with the percolation threshold of EP reported in the previous studies~\cite{Lee2011,Li2023}. It indicates that the percolation threshold in EP remains consistent regardless of competition rules governing the insertion of intra-cluster bonds. Furthermore, the critical exponent $\nu$ is also consistent across all three scenarios within the error margins, suggesting that the percolation transitions of CCs for different competition rules of intra-cluster bonds belong to the same universality class. It is worth noting that simulations in the conventional ensemble of fixed bond densities might erroneously suggest variations in percolation thresholds among different rules~\cite{Cho2011}.

Further, we can estimate the volume fractal dimension $d_f$ from the observable $C_1$ sampled at the dynamic pseudo-critical point, by fitting the data to the FSS ansatz
\begin{equation}
C_1=V^{d_f}(a_0+a_1V^{-\omega_1}+a_2V^{-\omega_2}+\cdots),    \label{Eq:PowerN}
\end{equation}
where $\omega_i (i=1,2)$ denotes the correction exponents. The stable fit can be obtained by including only one correction term ($a_2=0$) in Eq.~(\ref{Eq:PowerN}), and the results are listed in Tab.~\ref{t1}. The consistency of the fractal dimension across all three scenarios further emphasizes the independence of the EP nature from the competition rule of intra-cluster bonds, which is visually displayed by the nearly complete overlap of the FSS of $C_1$ for the three scenarios, as depicted in Fig.~\ref{f3} (c).

The discrepancy between the finite-size behaviors observed at the dynamic pseudo-critical point $\mathcal{T}_V$ and the infinite-volume critical point $T_c$, as depicted in Figs.~\ref{f2} and \ref{f3}, highlights an intriguing aspect of the EP dynamics. This contrast is further elucidated by studying the fluctuation $\sigma(\mathcal{T}_V)$ as a function of system volume $V$~\cite{Li2023}, as shown in Fig.~\ref{f3} (b). The well-defined scaling behavior $\sigma(\mathcal{T}_V)\sim V^{-\theta}$, where $\theta<1/\nu$, indicates that the pseudo-critical point of EP can deviate significantly from $T_c$ in some realizations, leading to distinct finite-size behaviors at $\mathcal{T}_V$ compared to $T_c$.

Moreover, it is proposed that the scaling window is effectively defined around $\mathcal{T}_V$ rather than $T_c$~\cite{Li2023}. Consequently, the FSS extracted at $T_c$ encompasses a mixture of behaviors observed over a wide range of bond densities, potentially spanning both super- and sub-critical phases. This mixture effect under different competition rules of intra-cluster bonds gives rise to the anomalous finite-size behaviors depicted in Fig.~\ref{f2}, highlighting the nuanced nature of the EP dynamics.

To determine the value of $\theta$, we fit the data of $\sigma(\mathcal{T}_V)$ to the scaling ansatz Eq.~(\ref{Eq:PowerN}), where the exponent $d_f$ is replaced by $-\theta$. Including one correction term, the stable fit results suggest a consistent exponent $\theta=1/2$ for various competition rules of intra-cluster bonds (Tab.~\ref{t1}). This finding supports the argument that the distribution of $\mathcal{T}_V$ in EP follows the central limit theorem and obeys a normal distribution~\cite{Li2023,Feshanjerdi2024}, which could be a universal property for EP of various rules.

\section{Percolation of biconnected clusters}    \label{sec-bc}

In this section, we study the percolation transition of BCs in EP under different competition rules of intra-cluster bonds. The data shown in this section is extracted at the dynamic pseudo-critical point $\mathcal{T}_{V}'$ identified by the largest one-step increment size of the largest BC.

\subsection{The asymptotic behavior of the pseudo-critical point $\mathcal{T}_{V}'$}

\begin{figure}
\centering
\includegraphics[width=\columnwidth]{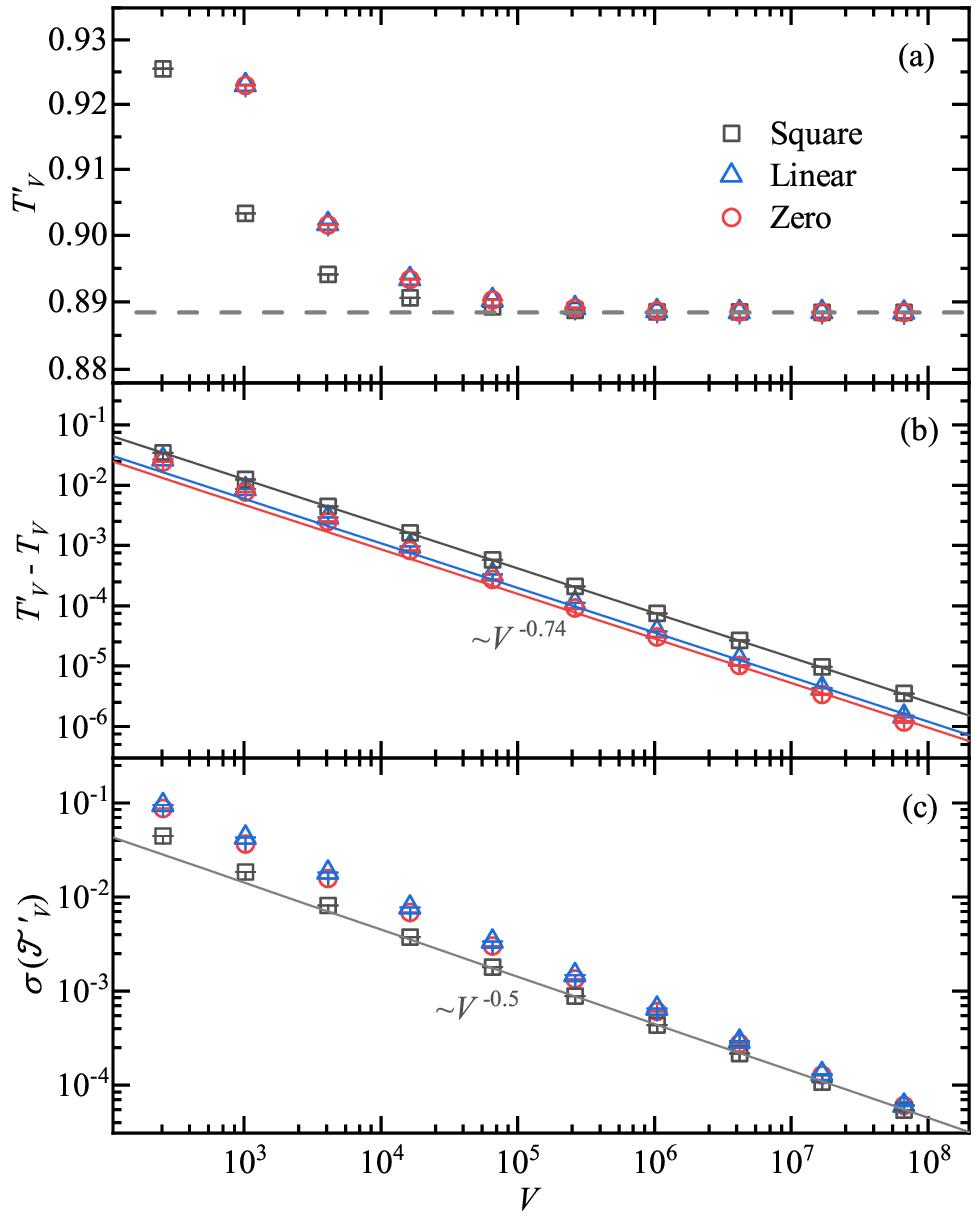}
\caption{The asymptotic behaviors of the pseudo-critical point $\mathcal{T}_{V}'$ identified by the one-step incremental size of the largest BC for different competition rules of intra-cluster bonds. (a) The pseudo-critical point $T_{V}'$ is plotted as a function of system volume $V$. Although for finite $V$, the three scenarios show different pseudo-critical points $T_{V}'$, they are convergent to the same percolation threshold $T_{V}'=0.888449$ for large systems, which is consistent with the percolation threshold of CCs. The fit results for these asymptotic behaviors are shown in Tab.~\ref{t2}. (b) The distance between the pseudo-critical points $T_{V}$ and $T_{V}'$ for different competition rules of intra-cluster bonds. All three lines represent the power-law decay with exponent $0.74$, which is just the reciprocal value of the correlation-length exponent $1/\nu$ listed in Tabs.~\ref{t1} and \ref{t2}. (c) The plots of the fluctuation of the dynamic pseudo-critical point $\sigma(\mathcal{T}_{V}')$ versus system volume $V$, indicating that all the three scenarios have the same scaling $\sigma \sim V^{-1/2}$.}  \label{f4}
\end{figure}

In Fig.~\ref{f4} (a), we plot the pseudo-critical point $T_{V}'$ versus the system volume $V$. It is evident that $T_{V}'$ varies for different competition rules of intra-cluster bonds. Because BCs are always formed out of CCs, a system with a large $T_{V}$ also has a large $T_{V}'$, as depicted in Fig.~\ref{f4} (a).

To capture the asymptotic behavior of the pseudo-critical point $T_{V}'$, we conduct a least-square fit to the Monte Carlo data of $T_{V}'$ using the scaling ansatz Eq.~\eqref{Eq:PowerC}. Accounting for systematic errors, we obtain estimates summarized in Tab.~\ref{t2}. The fit results reveal that the infinite-volume critical point coincides with the percolation threshold of CCs (Tab.~\ref{t1}), indicating that CCs and BCs percolate simultaneously in EP, irrespective of the competition rule of intra-cluster bonds.

\begin{table}[b]
\caption{The fit results of the infinite-volume critical point $T_{c}$, the reciprocal value of the correlation-length exponent $1/\nu$, the exponent $\theta$ to describe the fluctuation of $\mathcal{T}_{V}'$, and the fractal dimension $d_{f}'$ of the largest BC for different competition rules of intra-cluster bonds. Here, the data are all sampled at the dynamic pseudo-critical point $\mathcal{T}_{V}'$, where the one-step increment size of $\mathcal{B}_1$ reaches its maximum value. Within error bars, $T_{c}$, $1/\nu$, and $\theta$ take the same value as those obtained by CC (Tab.~\ref{t1}), which is independent of the competition rules of intra-cluster bonds. However, the fractal dimension $d_{f}'$ varies across different rules.}  \label{t2}
\begin{ruledtabular}
\begin{tabular}{llllll}
\text{Rule}   &  \multicolumn{1}{c}{$T_c$}  &  \multicolumn{1}{c}{$1/\nu$} &  \multicolumn{1}{c}{$\theta$} & \multicolumn{1}{c}{$d_{f}$} & \multicolumn{1}{c}{$d_{f}'$}  \\
\hline
\text{Square} & 0.888\,449\,4(3)     & 0.74(2)   & 0.49(2)  & 0.9347(2)  & 0.511(1)   \\
\text{Linear} & 0.888\,449\,3(5)     & 0.74(1)   & 0.50(9)  & 0.9363(5)  & 0.560(5)    \\
\text{Zero}   & 0.888\,449\,0(4)     & 0.74(1)   & 0.4(2)   & 0.9367(8)  & 0.575(8)
\end{tabular}
\end{ruledtabular}
\end{table}

Furthermore, within the margin of errors, the fit results in Tab.~\ref{t2} suggest identical exponents $1/\nu$ and $\theta$ as those for $\mathcal{T}_V$. This implies that the pseudo-critical points identified by CCs and BCs exhibit the same asymptotic behavior, corroborated by the power-law decay of $T_{V}'-T_{V}\sim V^{-1/\nu}$ depicted in Fig.~\ref{f4} (b). The scaling behavior of the fluctuation $\sigma(\mathcal{T}_V')\sim V^{-\theta}$ with $\theta=1/2$ is also evident in Fig.~\ref{f4} (c). Importantly, these scalings are independent of the competition rule of intra-cluster bonds, which solely influences finite-size corrections.

\subsection{Fractal dimension of biconnected clusters}

\begin{figure}
\centering
\includegraphics[width=\columnwidth]{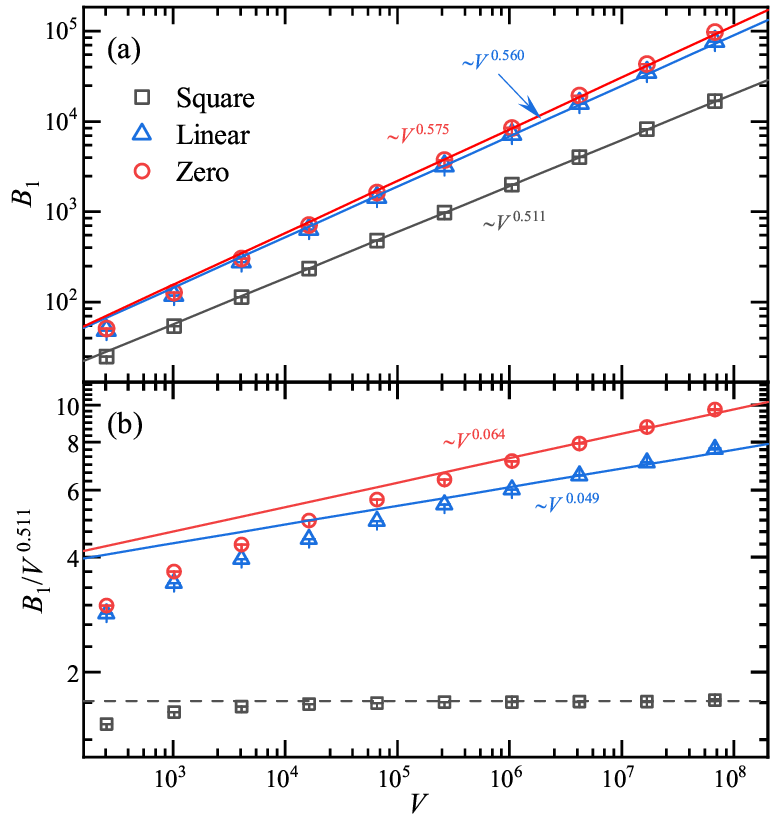}
\caption{The FSS of the size of the largest BC sampled at the dynamic pseudo-critical point $\mathcal{T}_{V}'$ for different competition rules of intra-cluster bonds. (a) The size $B_1$ of the largest BC as a function of system volume $V$. The fit results of Tab.~\ref{t2} are indicated by lines. (b) The ratio $B_1/V^{0.511}$ as a function of system volume $V$. For square rule, the fractal dimension is $d_{f}'=0.511$, thus, the ratio $B_1/V^{0.511}$ approaches a constant for large $V$, indicated by the dashed line. For the other two rules, the ratio $B_1/V^{0.511}$ shows a power-law growth for large $V$, and the exponent of the power-law growth is consistent with the fit result in Tab.~\ref{t2}. This suggests that the critical BC has different fractal dimensions under different competition rules of intra-cluster bonds.} \label{f5}
\end{figure}

In Fig.~\ref{f5} (a), we observe the power-law growth of the size $B_1$ of the largest BC sampled at the dynamic pseudo-critical point $\mathcal{T}_{V}'$ for different competition rules of intra-cluster bonds. This growth behavior signifies the fractal nature of the critical BC, with the fractal dimension $d_{f}'$ being dependent on the competition rule of intra-cluster bonds.

To quantify the fractal dimension $d_{f}'$ of the critical BC, we perform fits to the FSS ansatz Eq.~\eqref{Eq:PowerN} and summarize the estimates in Tab.~\ref{t2}. The differences between the $d_{f}'$ values obtained in different scenarios are significant, compared to the error bars, confirming distinct fractal dimensions of BCs. For better visualization of these differences, we plot the ratio $B_1/V^{0.511}$ for all three scenarios in Fig.~\ref{f5} (b). Notably, for square rule, where the fractal dimension is $d_{f}'=0.511$, the ratio $B_1/V^{0.511}$ tends to approach a constant for large $V$. However, for the other two rules, this ratio exhibits a power-law growth for large $V$, indicating different fractal dimensions of BCs.

For comparison, we also sample the size $C_1$ of the largest CC at the dynamic pseudo-critical point $\mathcal{T}_{V}'$. The fit results for $d_f$ are listed in Tab.~\ref{t2}, and within double error bars, these values are identical. This indicates that the fractal dimension $d_f$ of the critical CC at $\mathcal{T}_{V}'$ is independent of the competition rule of intra-cluster bonds and has the same value as the one sampled at $\mathcal{T}_{V}$ (Tab.~\ref{t1}). This consistency arises naturally as $\mathcal{T}_{V}$ and $\mathcal{T}_{V}'$ exhibit the same asymptotic behavior and is both situated within the scaling window $\mathcal{O}(V^{-1/\nu})$.

\begin{figure}
\centering
\includegraphics[width=\columnwidth]{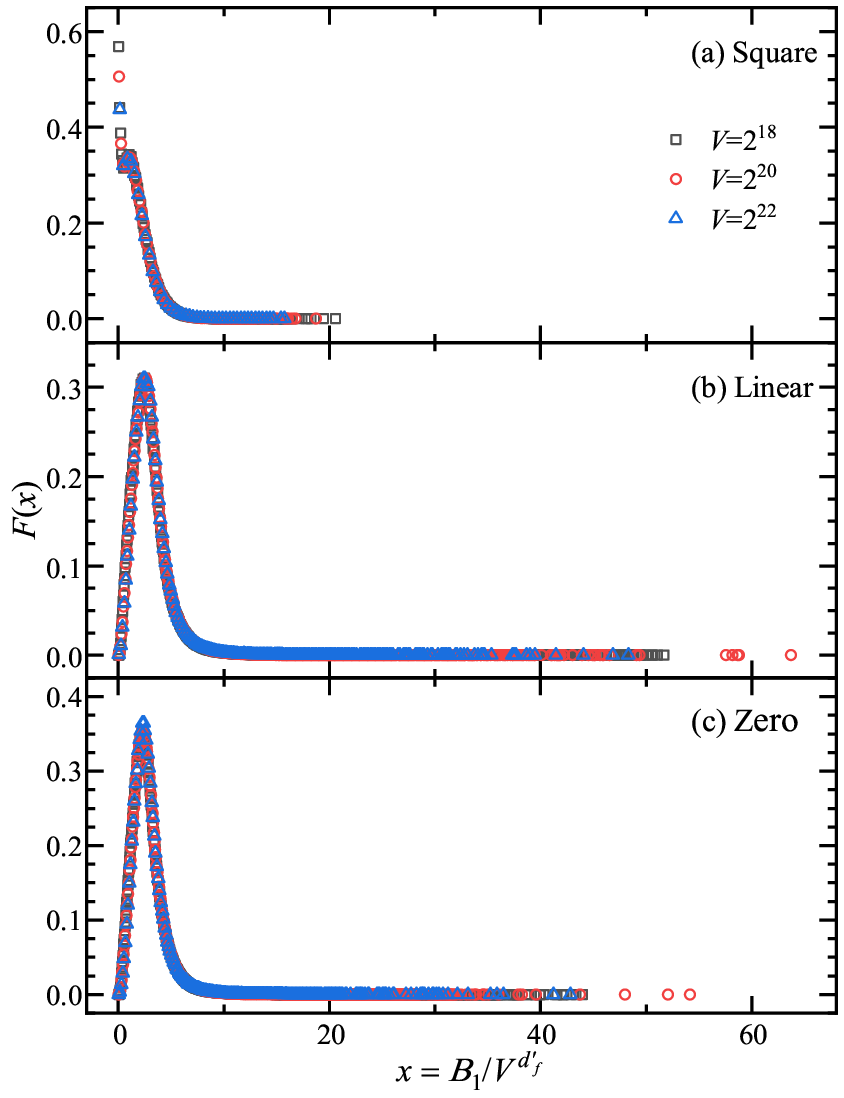}
\caption{The probability distribution $F(x)$ of the size of the largest BC sampled at the dynamic pseudo-critical point $\mathcal{T}_{V}'$ for square (a), linear (b), and zero (c) rules. By defining $x\equiv B_1/V^{d_{f}'}$ with the fit result of $d_{f}'$ in Tab.~\ref{t2}, all the three scenarios show nice collapses of data from different system volumes, confirming the unique $d_f'$ for each scenario.}   \label{f6}
\end{figure}

To further confirm the unique fractal dimension of the critical BC, we examine the probability distribution $F(x)$ of the size of the largest BC in Fig.~\ref{f6}. By defining $x\equiv B_1/V^{d_{f}'}$ using the fit result of $d_{f}'$ from Tab.~\ref{t2}, we achieve a well-renormalized distribution, demonstrating a collapse of data from different system volumes for all three scenarios. This validates the distinct fractal dimensions $d_{f}'$ for different competition rules of intra-cluster bonds.

\subsection{Cluster number density}

\begin{figure}
\centering
\includegraphics[width=\columnwidth]{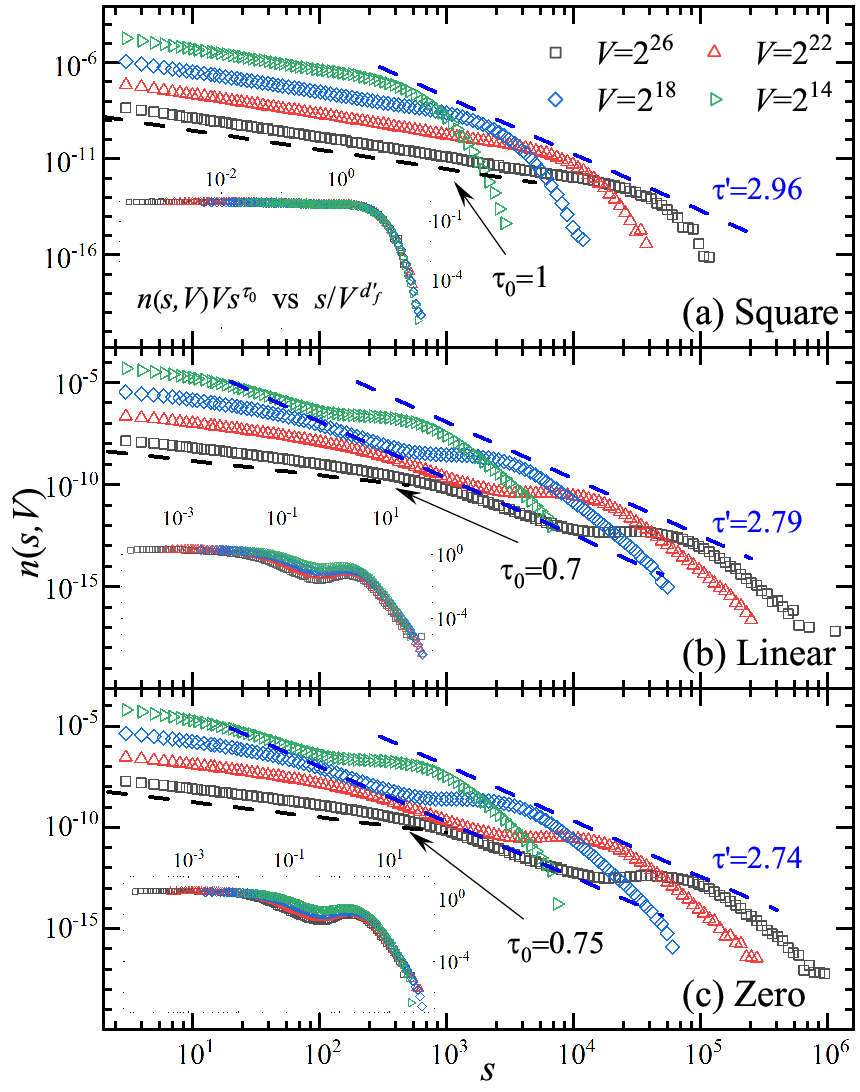}
\caption{The cluster number density of BCs for different competition rules of intra-cluster bonds. (a) Square rule. (b) Linear rule. (c) Zero rule. Two scalings can be observed for finite BCs, separated by a characteristic size $s_0$. The standard Fisher exponent $\tau'=1+1/d_f'$ dominates the size distribution for $s\gg s_0$, while the size distribution for $s\ll s_0$ shows a modified Fisher exponent $\tau_0$. The part $s<s_0$ decreases as a whole for increasing system volume $V$. The insets show the rescaled cluster number density of BCs $n(s,V)Vs^{\tau_0}$ as a function of $s/V^{d_{f}'}$, where the fit results of $d_{f}'$ in Tab.~\ref{t2} are used. The nice data collapse for $s<s_0$ suggests that $n(s,V)\sim V^{-1}$ for all the three scenarios. In addition, the plots also suggest $\tau_0\approx1$, $0.7$, and $0.75$ for the three scenarios, respectively.}    \label{f7}
\end{figure}

By utilizing the hyperscaling relation $\tau'=1+1/d_f'>2$ along with the fractal dimension $d_f'$ listed in Tab.~\ref{t2}, we can immediately determine the Fisher exponent $\tau'$ for BCs. However, the size distribution of BCs cannot be fully characterized by this standard Fisher exponent, instead, it exhibits a double-scaling behavior, as depicted in Fig.~\ref{f7}. Specifically, for linear and zero rules (Figs.~\ref{f7} (b) and (c)), apart from the standard characteristic size $s_\xi \sim V^{d_f'}$, another characteristic size $s_0$ emerges, which also grows as the system volume increases. For $s\ll s_0$, a modified Fisher exponent $\tau_0<\tau'$ is observed, while for $s\gg s_0$, the size distribution of BCs is predominantly governed by the standard Fisher exponent $\tau'$, rapidly decaying for $s>s_\xi$. Moreover, the overall cluster number density $n(s,V)$ for $s<s_0$ decreases with increasing system volume $V$. From this viewpoint of double scaling, systems of square rule correspond to an $s_0$ that is equal to or slightly smaller than $s_\xi$, resulting in a seemingly pure power-law distribution governed only by $\tau_0$, see Fig.~\ref{f7} (a).

From these observations, we propose an expression for the cluster number density $n(s,V)$ of BCs as follows:
\begin{equation}
n(s,V)=\left\{\begin{split}
                 & A s^{-\tau_0},  &s \ll s_0,   \\
                 &s^{-\tau'}\tilde{n}(s/s_\xi),  &s \gg s_0,
              \end{split}\right.        \label{eq-nsh}
\end{equation}
where $A$ is a $V$-dependent parameter ensuring the normalizability of the cluster size distribution $sn(s,V)$ for $\tau_0\leq2$. The normalizing condition $A\int_{1}^{s_0} s^{1-\tau_0}ds + \int_{s_0}^{s_\xi} s^{1-\tau'}ds\sim \mathcal{O}(1)$ yields $A\leq\mathcal{O}(s_0^{\tau_0-2})$, where $\tau_0<2$ and $\tau'>2$ from the observation in Fig.~\ref{f7}. The insets of Fig.~\ref{f7} demonstrate $n(s,V)Vs^{\tau_0}$ as a function of $s/V^{d_f'}$, with data from various system volumes collapsing well. This collapse indicates $A\sim V^{-1}$ for all three scenarios, regardless of $\tau_0$. It is worth noting that for standard percolation on random graphs, $n(s,V)\sim V^{-1}$ for BCs as well~\cite{Ben-Naim2005,Huang2018}.

\begin{figure}
\centering
\includegraphics[width=\columnwidth]{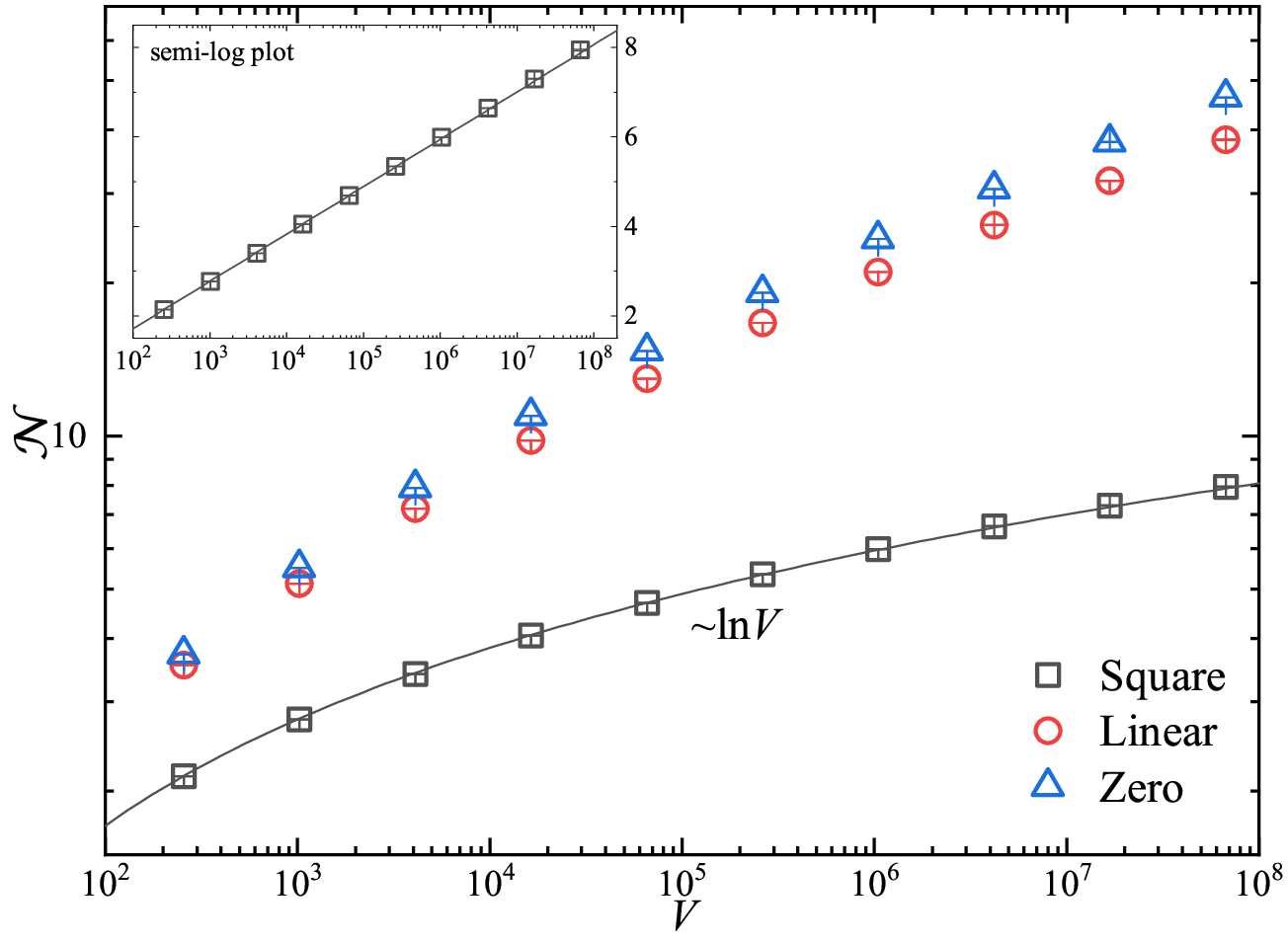}
\caption{The total number of BCs $\mathcal{N}$ at the dynamic pseudo-critical point $\mathcal{T}_{V}'$ for different competition rules of intra-cluster bonds. For square rule, $\mathcal{N}$ presents a logarithmic growth with increasing system volume $V$. The line shows the function of $\mathcal{N}\sim\ln V$. The straight line of the same data in the semi-log plot also confirms the logarithmic growth of $\mathcal{N}$, see the inset. For linear and zero rules, the growth of $\mathcal{N}$ can neither be fitted by a power law function nor a logarithmic function, which would have some intricate finite-size corrections that are not represented by the used fit function.}    \label{f8}
\end{figure}

To further understand the different $n(s,V)$ of the three scenarios in Fig.~\ref{f7}, we study the FSS of the total number of BCs, calculated as $\mathcal{N}=V\sum_{s=1}n(s,V)$. Calling Eq.~(\ref{eq-nsh}), the total number of BCs can be estimated as
\begin{equation}
\mathcal{N}\sim A V\int_{1}^{s_0}s^{-\tau_0}ds + V\int_{s_0}^{s_\xi}s^{-\tau'}ds.   \label{eq-nt}
\end{equation}
The FSS behavior of $\mathcal{N}$ is dependent on $s_0$ and $\tau_0$. If $s_0$ is absent, Eq.~(\ref{eq-nt}) yields $\mathcal{N}\sim V$, which corresponds to the observation for CCs. However, the simulation results in Fig.~\ref{f8} clearly demonstrate that for all three scenarios, $\mathcal{N}$ of BCs diverges slower than $\sim V$, suggesting a non-trivial $s_0$. For square rule, the total number of BCs is well described by the logarithmic function, i.e., $\mathcal{N}\sim \ln V$. To account for this behavior using Eq.~(\ref{eq-nt}), it requires $\tau_0=1$ in the first term, and $s_0\sim s_\xi$ in the second term. This explains the scaling behavior in Fig.~\ref{f7} (a), where all finite BCs exhibit a size distribution with $\tau_0=1$.

The data collapse in the insets of Figs.~\ref{f7} (b) and (c) suggest $\tau_0\approx0.7$ and $0.75$ for linear and zero rules, respectively. For these $\tau_0<1$, both terms in Eq.~(\ref{eq-nt}) diverge as $s_0\to\infty$ for $V\to\infty$, and the leading behavior depends on the FSS behavior of $s_0$. Due to the lack of direct measurement for $s_0$, its FSS is unavailable in our phenomenological discussion. From Fig.~\ref{f8}, where the divergence of $\mathcal{N}$ for linear and zero rules cannot be captured by a simple logarithmic or power-law function, it is suggested that the FSS of $s_0$ should include strong finite-size corrections.

From the preceding discussion, we ascertain that the modified Fisher exponent $\tau_0$ stems from the vanishing cluster number density $n(s,V)$ for $s<s_0$. Such vanishing cluster number density phenomena have also been observed for leaf-free and bridge-free clusters in high-dimensional percolation~\cite{Huang2018} and holes in no-enclave percolation~\cite{Hu2016}. Here, the cluster number density of BCs might exhibit a more intricate behavior, contingent upon the competition rule of intra-cluster bonds.

\section{Conclusion}

In this study, we delve into the percolation transition of high-order connectivity in EP through three specific competition rules of intra-cluster bonds in the Achlioptas process. Extensive simulations corroborate that EP, regardless of competition rules applied to intra-cluster bonds, conforms to the same percolation threshold and universality class. This clarifies that the competition rules of intra-cluster bonds do not affect the critical behaviors of EP. However, the finite-size behaviors of BCs are very sensitive to these rules, and we provide strong numerical evidence demonstrating the rule-dependent fractal dimensions of BCs. Additionally, BCs exhibit unique properties, such as a double-scaling behavior in size distribution, requiring a modified Fisher exponent to describe the size distribution of small BCs.

Our findings contribute to resolving the debate regarding the universality of EP in relation to the competition rules of intra-cluster bonds, and demonstrate the superiority of the event-based ensemble over the conventional fixed bond density ensemble in accurately extracting the FSS behavior. Building on these findings, we reveal the presence of non-trivial high-order connectivity within percolation clusters, despite EP focusing solely on simply connected clusters. Therefore, it would be intriguing to explore the emergence of other high-order connectivities in EP and investigate the potential existence of a genuine discontinuous transition within the Achlioptas process.

\section*{Acknowledgments}

The authors acknowledge helpful discussions with Sheng Fang. The research was supported by the Fundamental Research Funds for the Central Universities (No.~JZ2023HGTB0220).

\bibliography{ref}

\end{document}